\newcommand{\changes}[1]{\textcolor{black}{#1}}
\documentclass[sigconf,authorversion,nonacm]{acmart}
\AtBeginDocument{%
  \providecommand\BibTeX{{%
    \normalfont B\kern-0.5em{\scshape i\kern-0.25em b}\kern-0.8em\TeX}}}

\begin{document}

%%
%% The "title" command has an optional parameter,
%% allowing the author to define a "short title" to be used in page headers.
\title{UICrit: Enhancing Automated Design Evaluation with a UI Critique Dataset}

%%
%% The "author" command and its associated commands are used to define
%% the authors and their affiliations.
%% Of note is the shared affiliation of the first two authors, and the
%% "authornote" and "authornotemark" commands
%% used to denote shared contribution to the research.
\author{Peitong Duan}
\email{peitongd@berkeley.edu}
\authornote{This work was completed while the author was an intern at Google.}
\affiliation{%
  \institution{UC Berkeley}
  \city{Berkeley}
  \country{CA, USA}
}

\author{Chin-Yi Chen}
\email{cchinyi@google.com}
\affiliation{%
  \institution{Google DeepMind}
  \city{Mountain View}
  \country{CA, USA}
}

\author{Gang Li}
\email{leebird@google.com}
\affiliation{%
  \institution{Google DeepMind}
  \city{Mountain View}
  \country{CA, USA}
}

\author{Bjoern Hartmann}
\email{bjoern@eecs.berkeley.edu}
\affiliation{%
  \institution{UC Berkeley}
  \city{Berkeley}
  \country{CA, USA}
}

\author{Yang Li}
\email{liyang@google.com}
\affiliation{%
  \institution{Google DeepMind}
  \city{Mountain View}
  \country{CA, USA}
}

%%
%% By default, the full list of authors will be used in the page
%% headers. Often, this list is too long, and will overlap
%% other information printed in the page headers. This command allows
%% the author to define a more concise list
%% of authors' names for this purpose.
\renewcommand{\shortauthors}{Peitong Duan, et al.}

%%
%% The abstract is a short summary of the work to be presented in the
%% article.
\begin{abstract}
Automated UI evaluation can be beneficial for the design process; for example, to compare different UI designs, or conduct automated heuristic evaluation. LLM-based UI evaluation, in particular, holds the promise of generalizability to a wide variety of UI types and evaluation tasks. However, current LLM-based techniques do not yet match the performance of human evaluators. We hypothesize that automatic evaluation can be improved by collecting a targeted UI feedback dataset and then using this dataset to enhance the performance of general-purpose LLMs. We present a targeted dataset of 3,059 design critiques and quality ratings for 983 mobile UIs, collected from seven experienced designers. We carried out an in-depth analysis to characterize the dataset's features. We then applied this dataset to achieve a 55\% performance gain in LLM-generated UI feedback via various few-shot and visual prompting techniques. We also discuss future applications of this dataset, including training a reward model for generative UI techniques, and fine-tuning a tool-agnostic multi-modal LLM that automates UI evaluation.
\end{abstract}

%%
%% The code below is generated by the tool at http://dl.acm.org/ccs.cfm.
%% Please copy and paste the code instead of the example below.
%%


%%
%% Keywords. The author(s) should pick words that accurately describe
%% the work being presented. Separate the keywords with commas.

%% A "teaser" image appears between the author and affiliation
%% information and the body of the document, and typically spans the
%% page.

\received{3 April 2024}
%\received[revised]{12 March 2009}
%\received[accepted]{5 June 2009}

%%
%% This command processes the author and affiliation and title
%% information and builds the first part of the formatted document.
\maketitle

\section{Introduction}
Feedback is essential for guiding designers towards improving their user interface design. However, human feedback is not always readily available. While automated UI evaluation methods are able to provide instantaneous feedback, they have other limitations. Non-LLM based methods have limited generalizability and require large amounts of training data in order to accomplish a specific task, such as predicting user task completion time \cite{10.1145/3313831.3376589}. While LLM-based methods are able to generalize, their performance for UI feedback generation has potential for improvement \cite{duan2024}.

Some of the shortcomings of prior LLM-based UI evaluation can be attributed to limitations of LLMs at the time, such as lack of multi-modal input and small context windows. However, other weaknesses may be due to more fundamental data gaps in the performance of LLMs for UI evaluation (pointed out by \cite{duan2024}), which include poor knowledge of popular design conventions and failure to properly prioritize in cases when design guidelines clash. These limitations could be potentially addressed by few-shot training or fine-tuning on a ground truth dataset of high quality design feedback provided by human experts, where the LLM could learn the specifics and nuances of UI evaluation. However, the research community currently lacks a large enough dataset to comprehensively capture the knowledge required to carry out high quality UI evaluation.  

To advance the effort towards improving automated UI feedback, we introduce UICrit, the first targeted dataset of design critiques for 983 mobile UI screens, consisting of 3,059 natural language design critiques collected from seven experienced designers. To help contextualize the feedback, each critique contains a bounding box highlighting relevant regions in the UI screenshot, and the dataset also includes numerical ratings for the aesthetics, usability, and overall design quality for each UI screen, which were manually determined from its design critiques and a carefully constructed rubric. We then analyzed this dataset and categorized the topics covered by the design critiques into five broad categories (layout, color contrast, text readability, button usability, learnability), and also determined the types of UIs represented in this dataset and the distribution of critiques targeting individual UI elements, groups of elements, and the entire screen. We also identified underrepresented design issues and UI types in this dataset, so future work could address these gaps.

We demonstrate this dataset's utility by applying it to automate two feedback generation tasks that are useful to designers: obtaining feedback for a region of interest in a UI screen and attaining feedback, corresponding bounding boxes, and numerical design quality ratings for the entire UI screen. We leverage the latest advancements in LLMs to automate these tasks with \textit{only} the UI screenshot as input and utilizing few shot prompting, with samples taken from the dataset. We experimented with various few-shot and visual prompting techniques to improve the quality of the generated comments, scores, and corresponding bounding boxes. We found that our best prompt design involves 1) using few-shot examples with similar task and visual embedding to the input UI screen, and 2) enhancing the screenshot by displaying coordinate references along its edge to help with bounding box estimation. This prompt setup resulted in a 55 percent improvement in performance over zero shot prompting, which was confirmed by design experts via a user study, demonstrating the value of this dataset in improving automated design comments generation. 

Finally, we discuss potential broader applications of this dataset. These applications include training a reward model to predict design critiques and numerical design quality scores for the output of generative UI models, and fine-tuning a tool-agnostic LLM that generates high quality design feedback given only the UI screenshot, which makes it amenable to be integrated into any design tool to automate mockup evaluation.

In summary, our contributions are as follows:
\begin{itemize}
    \item A dataset with 3,059 design critiques, corresponding boxes to mark relevant regions, and design quality ratings for 983 distinct UI screens. This dataset was collected from experienced designers through a carefully designed protocol that ensures accurate design feedback and ratings. The dataset is available on  github\footnote{\changes{\url{https://github.com/google-research-datasets/uicrit}}}
    \item An in-depth analysis of the dataset to understand how professional designers evaluate a design, and extract informative features of this dataset, such as the types of design issues covered by the critiques, types of UIs represented, and the distribution of element, group, and screen-level comments. We also identify a few underrepresented design issues and UI types in this dataset, so future data collection efforts could address these gaps.  
    \item A novel prompt chain design (illustrated in Figure \ref{fig:promptchain}) that queries a multi-modal LLM with the UI screenshot and receives design feedback, corresponding bounding boxes, and design quality ratings. The first prompt queries the LLM for design feedback and ratings with few-shot samples selected from the dataset based on task and visual similarity to the input UI. The second prompt takes the generated critiques and and makes an LLM call to determine the bounding boxes for each critique with visual prompting. We verified that this prompt design generated design feedback that was significantly better than zero shot prompting via a user study with design experts.
\end{itemize}
\section{Related Work}
\subsection{UI Datasets}
A myriad of UI-related datasets have been developed. There are UI datasets consisting of screenshots and XML representations \cite{Deka:2017:Rico, DBLP:journals/corr/abs-1910-10683, 10.1145/3544548.3581158}, as well as datasets augmented with human usage data, such as gaze patterns \cite{jiang2023ueyes, 10.1145/3379503.3403557, 9779948, shen2014webpage} and task traces \cite{rawles2023android, burns2022motifvln, DBLP:journals/corr/abs-2005-03776}. The RICO dataset \cite{Deka:2017:Rico} was one of the first large-scale UI datasets, which contains over 66k mobile UI screenshots and view hierarchies from 9.3k Android apps. RICO also contains other metadata, such as the traces from random exploration of the apps and UI layout vectors. Deka et. al. collected the RICO dataset via a system that combined both crowdsourcing and an automated mining system to further explore UI states. There have been several studies that refined the data in RICO\cite{clay, 10.1145/3406324.3410710}. Li et. al. created the CLAY dataset \cite{clay}, which is a cleaned subset the RICO dataset, where UIs with invalid view hierarchy layouts were removed. The authors then used the CLAY dataset to train deep learning models to automatically denoise mobile UI layouts.

There are also UI datasets that are augmented with human usage data. For instance, Jiang et. al. collected the UIEyes dataset \cite{jiang2023ueyes}, which consists of gaze data (fixation points and scanpaths) from 62 participants on 1,980 UIs collected from a large eye-tracking study. The authors then analyzed the dataset to determine the effects of factors like location and color on gaze behavior. Burns et. al. created the MoTIF dataset \cite{burns2022motifvln}, which contains descriptions of high level tasks to complete on the app, interaction traces of humans attempting to complete the tasks, and feasibility annotations of whether or not the task could be completed. The dataset was collected through a multi-step process where human workers annotated potential tasks, attempted to complete the tasks, and marked which ones were feasible. The authors then trained a model to predict task feasibility given the task and the interaction trace of the task attempt. However, despite these numerous UI datasets with human data, there currently does not exist a dataset of UIs with expert-annotated design critiques and design quality ratings.
\subsection{Automated UI Evaluation}
Prior to LLMs, automated methods to evaluate UIs include metrics \cite{10.1145/3266037.3266087, lee2020guicomp, 10.1145/2901790.2901817} and models that predict user behaviors, such as task completion time \cite{10.1145/3313831.3376589}, gaze patterns \cite{lee2020guicomp, 10.1145/3379337.3415825}, and user engagement \cite{10.1145/3313831.3376324}, which provide feedback that designers could use to revise their designs. Oulasvirta et. al. created the Aalto Interface Metrics \cite{10.1145/3266037.3266087}, which is a set of 17 metrics collected from prior studies and includes metrics like visual search performance and visual clutter. They then built a website, where users could upload their designs to be evaluated by these metrics. Wu et. al. collected a large dataset of human annotated ratings for the engagment of UI animations, and used it to train a neural network to predict the engagement ratings of an input UI animation \cite{10.1145/3313831.3376324}. The authors also developed a web app that uses this model to predict user engagement, and if the predict user engagement was low, it returns a set of potential reasons from a pre-defined set. However, these pre-LLM automated methods are unable to generalize beyond the specific design aspect they were developed to evaluate. Furthermore, they also require significant effort to achieve high performance; analytical metrics require considerable manual effort to develop, and data driven methods require vast amounts of training data. 

With the emergence of LLMs, Duan et. al. assessed their performance in automating UI evaluation \cite{duan2024}. Specifically they built a Figma plugin that could automatically carry out general-purpose heuristic evaluation for any Figma mockup and any arbitrary set of heuristics by querying GPT-4 with a JSON representation of the Figma mockup and the heuristics text, without any training data. They then carried out user studies to determine the performance and qualitative strengths and limitations of GPT-4 for heuristic evaluation, as well as how this LLM-based tool could fit into existing design practices. However, there were several limitations with their approach. GPT-4 at the time could only accept text-input, so they were limited to an XML-based representation of the UI mockup, which impedes the LLM's visual understand of the UI. Furthermore, XML-based representations are long, and given the shorter GPT-4 context windows at the time, they were limited to zero-shot prompting, which likely caused some of its performance issues. We extend their approach by leveraging Gemini Pro Vision, a recently launched multi-modal LLM, to automatically provide UI design feedback with the screenshot as input. We further collect a dataset of natural language UI critiques and ratings for 983 UI screens that could be used to improve LLM generated feedback via few-shot prompting and fine-tuning. We then utilized this dataset to explore various few-shot and visual prompting techniques, and found a configuration that considerably outperformed zero-shot prompting. 
\subsection{Design Feedback Support: Frameworks and Guidelines}
Accurate evaluation of a UI design requires a complex, multi-faceted approach \cite{articlerubric}. For instance, Hartmann et. al. developed a framework for evaluating UI design quality that separated it into five criteria: usability, content, aesthetics, reputation, and customization. These individual criteria are still quite broad; for example, Norman et. al. defined usability by five quality components: learnability, efficiency, memorability, user errors, and satisfaction \cite{Nielsen2024}. Furthermore, guidelines and heuristics have been developed to assist in UI evaluation. These guidelines and heuristics contain specific rules that good design should follow, and they are used in methods like heuristic evaluation \cite{Nielsen1990HeuristicEO}, where an evaluator identifies heuristic violations in a given design. A number of different sets of heuristics and guidelines have been developed, such as the Apple Human Interface Guidelines, which contain guidance and best practices for general UI design \cite{apple_hig}, as well as heuristics targeted to specific aspects of the design, such as Nielsen Norman's 10 Usability Heuristics \cite{10.1145/191666.191729}, the CrowdCrit Visual Design Critiques \cite{luther2015structuring}, and a set of guidelines for designing semantically coherent UIs \cite{duan2023towards}. Our work utilizes design guidelines and evaluation frameworks during the data collection to ensure that design critiques are grounded in best design practices and that the ratings accurately reflect the design quality.
\section{Data Collection}\label{sec:datacollection}
The goal of this data collection was to obtain a large dataset consisting of UI screenshots with corresponding design feedback, bounding boxes of screen regions being critiqued, and design quality ratings. \changes{We recruited seven annotators with prior professional design experience from a contracting company. Table \ref{tab:annotaterdetails} details the areas of design expertise and number of years of professional design experience for each annotator.} This section describes the annotation process. 
\begin{table}[t]
\centering
\begin{tabular}{|c|c|c|}
\toprule
\textbf{Annotator} & \textbf{Expertise} & \textbf{Years of Experience}\\ \midrule
\changes{A1} & \changes{UI/UX, Web} & \changes{1} \\ \hline
\changes{A2} & \changes{UI/UX, Web} & \changes{1}\\ \hline
\changes{A3} & \changes{UI/UX} & \changes{1.5} \\ \hline
\changes{A4} & \changes{UI/UX, Graphic} & \changes{1.8}\\ \hline
\changes{A5} & \changes{UI/UX, Graphic} & \changes{2}\\ \hline
\changes{A6} & \changes{UI/UX, Visual, UX Research} & \changes{1}\\ \hline
\changes{A7} & \changes{UI/UX, Graphic} & \changes{16}\\ \hline
\bottomrule
\end{tabular}
\caption{\changes{The areas of design expertise and number of years of professional design experience for each of the 7 dataset annotators.}}
\label{tab:annotaterdetails}
\end{table}
\subsection{Method}
\begin{figure}[h]
  \centering
  \includegraphics[width=\linewidth]{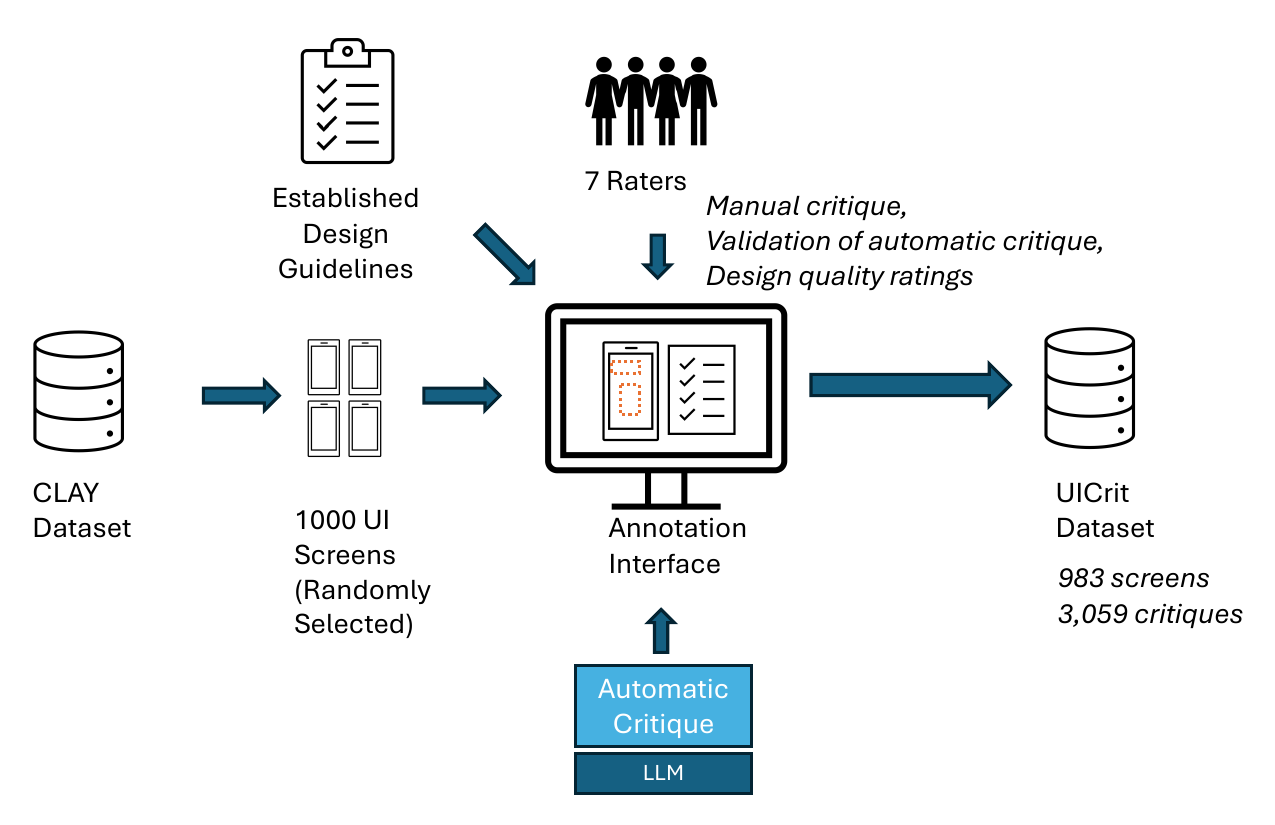}
  \caption{An illustration of the data collection process to obtain design comments, bounding boxes of critiqued screen regions, and design quality scores for 1000 UI screens.}
  \Description{A woman and a girl in white dresses sit in an open car.}
  \label{fig:ratingprocess}
\end{figure}
\begin{figure*}
  \centering
  \includegraphics[width=\linewidth]{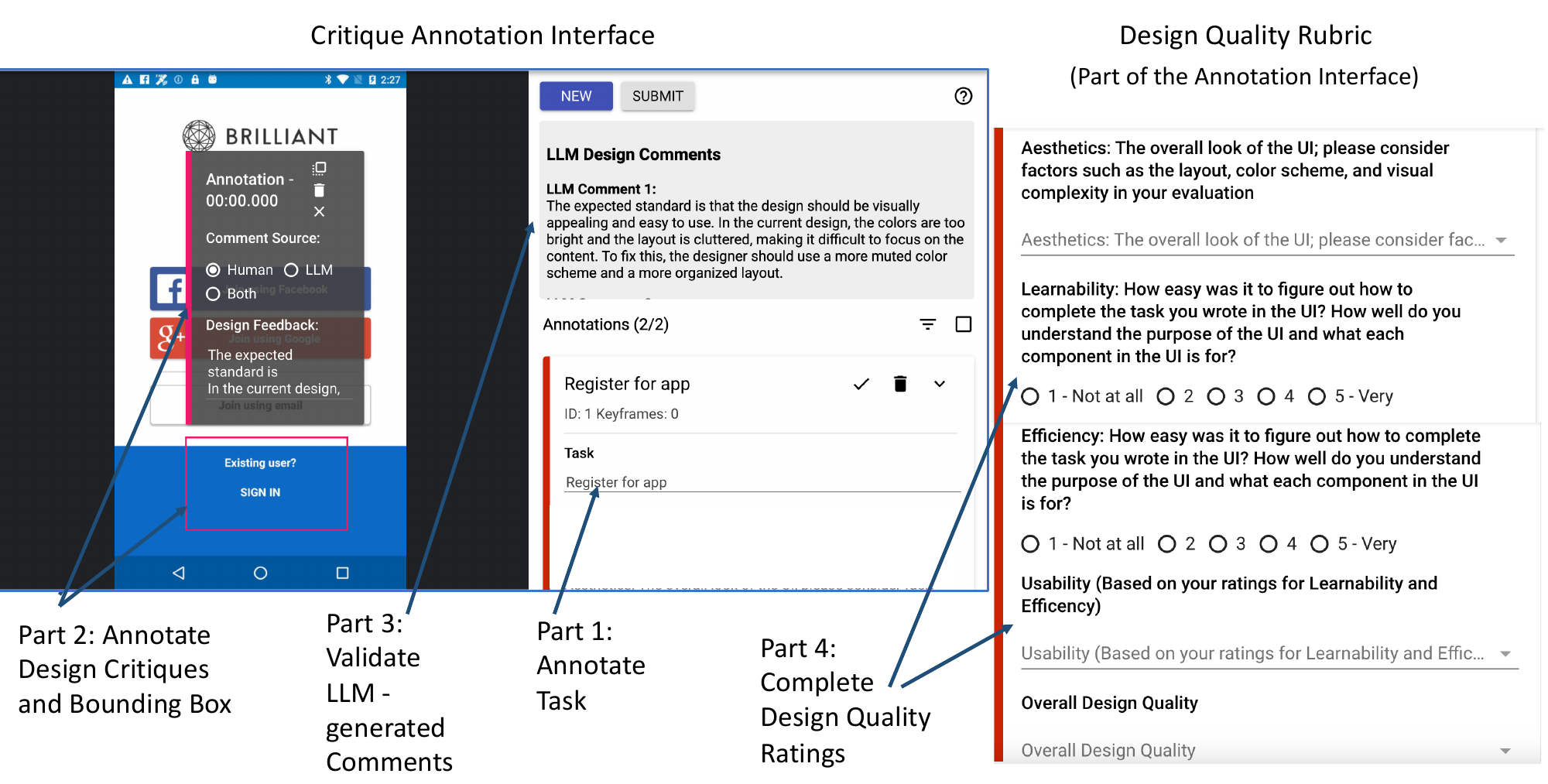}
  \caption{The interface for design critique and rating annotation, with regions corresponding to each part of the annotation marked.}
  \Description{A woman and a girl in white dresses sit in an open car.}
  \label{fig:datacollectionui}
\end{figure*}
Figure \ref{fig:ratingprocess} illustrates the annotation process. We implemented an interface on top of an existing internal crowdsourcing infrastructure, which assigns annotation tasks to workers. Before starting the data collection, we held an orientation session with the participants where we went through instructions and expectations. To ensure workers provide high quality design comments that are grounded in existing best practices, we provided three well-established design guidelines for them to reference during the annotation: the Nielsen Norman 10 Usability Heuristics \cite{10.1145/191666.191729}, CrowdCrit Visual Design Critiques \cite{luther2015structuring}, and the Apple Human Interface Guidelines \cite{apple_hig}. We instructed the workers to provide critiques based on these guidelines, as well as additional critiques drawn from their prior design experience, since UI design often goes beyond following guidelines. Furthermore, following \cite{duan2024}, we asked workers to follow Sadler's \cite{sadler1989formative} format for effective feedback and include these three things in each design critique: the expected standard (i.e. what good design should look like), the gap between the current design and the expected standard (i.e. the design issue), and how to close the gap (i.e. how to fix the current design). 

Each worker annotates critiques and ratings for a single UI screen at a time. The worker performs all the annotations for the UI on a single page, as shown in Figure \ref{fig:datacollectionui}. The annotation process is divided into four parts. Part 1 consists of inspecting the UI screen and writing down the main task supported by the interface (e.g. "register for app" for the screen in Figure \ref{fig:datacollectionui}). In addition to contributing valuable metadata to the dataset, asking participants to record the task ensures they thoroughly understand the UI before carrying out the evaluation, and the task also provides helpful context for evaluating the UI's usability at a later part of the annotation. Part 2 involves workers providing their own design critiques and corresponding bounding boxes. 
%To add a comment, they would first draw the bounding box, fill in text field with the critique text, and select "Human" as the comment source. 
Part 3 entails filtering for valid LLM comments. To supplement the critiques provided by each worker, we also pre-generated LLM design comments for each UI screen. However, LLMs may generate invalid design feedback, as found by \cite{duan2024}, so workers were instructed to read through each LLM-generated design comment and determine which ones are valid. 
%They would follow the process in Part 2 to annotate each valid comment as a design critique, except they would select "LLM" as the comment source or "Both" if the LLM comment is a repeat of a design critique they added in Part 1. 
Since LLMs have poor object localization \cite{dorkenwald2024pin}, only the LLM's comment text is provided, and the worker would have to determine the appropriate bounding box for valid LLM comments. Worker would also make minor edits to the text of valid LLM comments, if necessary, and also mark LLM comments that overlapped with ones they provided in Part 1.

Finally, in Part 4, workers provided the design quality ratings. To ensure accurate ratings, we designed a rubric that break downs the design quality evaluation into specific factors that are easier to assess. Figure \ref{fig:datacollectionui} shows the entire rubric, which is part of the critique annotation interface. The rubric first breaks down design quality into aesthetics and usability, which is modeled from the framework developed by \cite{articlerubric} that separates design evaluation into multiple criteria. We used the two criteria that are relevant to single screen UIs. To facilitate the rating process, we provide definitions for each criterion, factors to consider, and reminded participants to consider relevant critiques they provided, as illustrated in Figure \ref{fig:datacollectionui}. We further broke down usability into learnability and efficiency, which are two relevant factors to consider, according to \cite{Nielsen2024}. Since learnability and efficency are more specific than other dimensions in the rubric, they were rated on a 5-point Likert scale. Other dimensions (including the overall design quality) were rated on a 10-point scale. Participants were instructed to fill out the rubric and then come up with an overall rating of the design quality based on their ratings for each dimension.

We optimized for coverage, and only assigned one worker per UI to ensure we obtain annotations for a large set of UIs. We randomly selected a total of 1000 UIs to be annotated, taken from the CLAY dataset \cite{clay}. We used random selection to maximize the generalizability of this set of UIs. We then prompted Gemini Pro Vision \cite{team2023gemini} (zero-shot) with each UI screenshot and the text of the three guidelines used for this data collection to obtain the pre-generated LLM design critiques. 

\subsection{Results}
\begin{figure*}
  \centering
  \includegraphics[width=\linewidth]{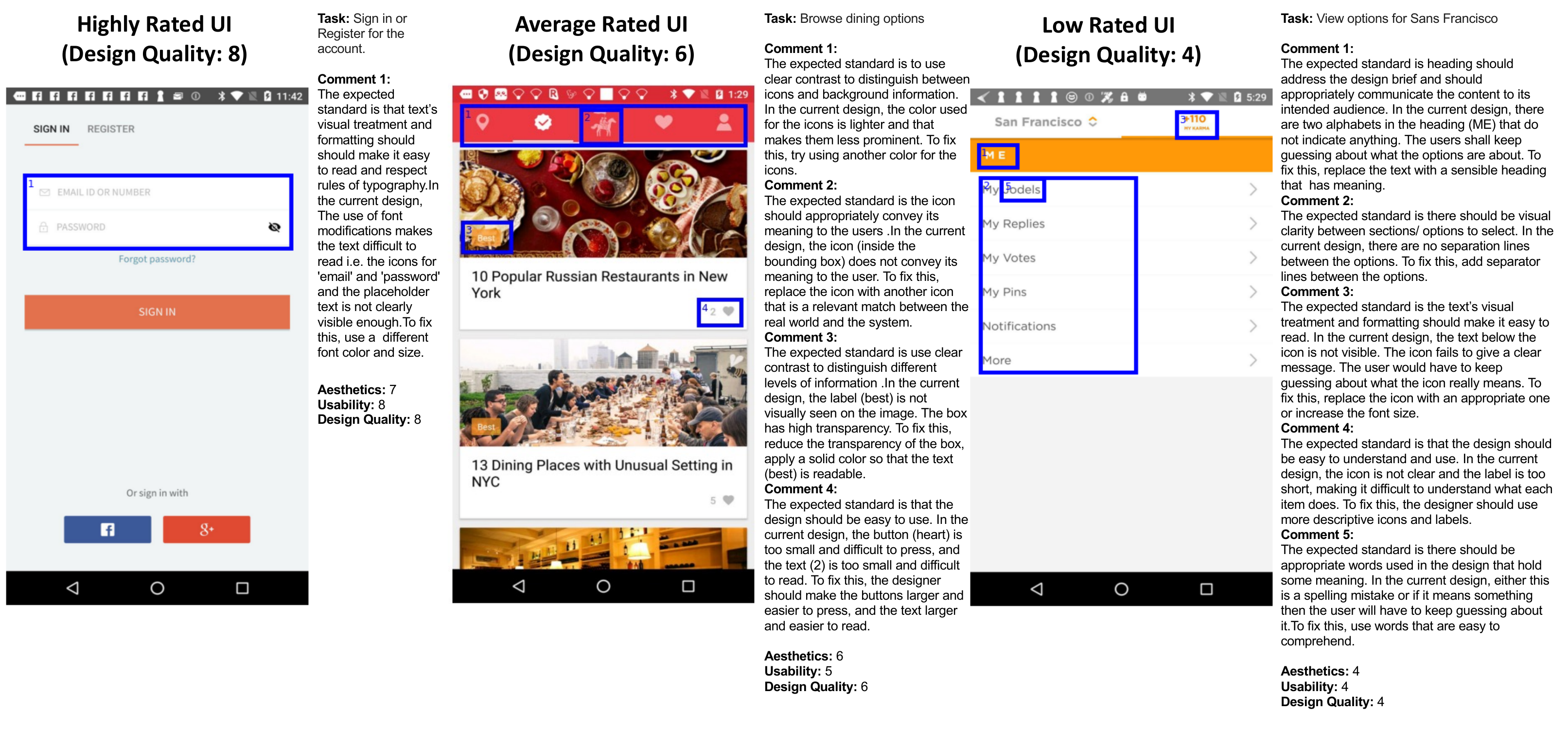}
  \caption{Examples of the data provided for each UI in the dataset. This figure shows worker annotations for a highly-rated UI, an average-rated UI, and a low-rated UI.}
  \Description{A woman and a girl in white dresses sit in an open car.}
  \label{fig:raterannotationexamples}
\end{figure*}
The data collection took around two weeks to complete, with all seven annotators working full-time. At the conclusion of the data collection, we obtained clean annotations (i.e. not missing any data) for 983 UI screens and collected a total of 3,059 critiques. The data for each UI screen includes the RICO ID, which can be used to access the screenshot, android view hierarchy, and other metadata from the original RICO dataset, the main task the UI screen is designed for, a set of design critiques with bounding box coordinates of corresponding screen regions, and numerical ratings along various dimensions including the aesthetics, usability, and overall design quality. Figure \ref{fig:raterannotationexamples} illustrates examples of this data for a UI with low design quality, a UI with average quality, and a UI with high design quality. Section \ref{sec:analysis} contains additional details and analyses of this dataset.
 
\section{Dataset Analysis}\label{sec:analysis}
\begin{figure}[h]
  \centering
  \includegraphics[width=\linewidth]{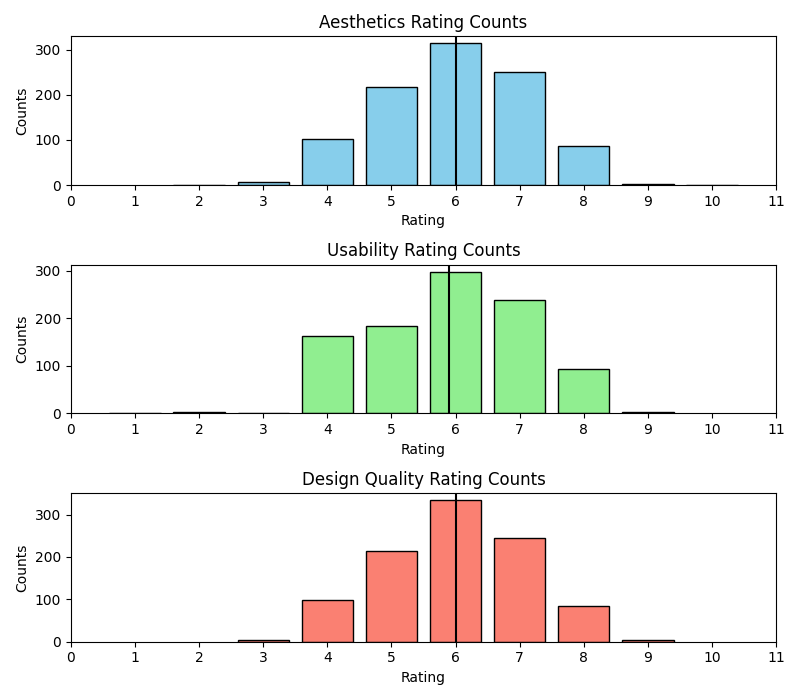}
  \caption{Histograms showing the counts for each numerical rating for aesthetics, usability, and overall design quality. The ratings all generally follow a normal distribution.}
  \Description{A woman and a girl in white dresses sit in an open car.}
  \label{fig:designratinghistogram}
\end{figure}
To better understand the dataset, we first compute high-level quantitative metrics. As stated earlier, we collected a total of 3,059 critiques for 983 UIs, which means each UI has on average, 3 critiques. Out of the 3,059 critiques, 2283 (74.6 percent) came from human designers, 256 (8.3 percent) came from Gemini, and 520 (17.1 percent) were provided by both humans and Gemini. Furthermore, Gemini generated a total of 5927 comments for this set of UIs, and only 776 were validated, which meant only 13.1 percent of the design comments generated by Gemini were valid. This indicates few-shot prompting or finetuning is needed for general purpose LLMs to perform the critique task effectively.  

Figure \ref{fig:designratinghistogram} contains histograms depicting the ratings count for aesthetics, usability, and the overall design quality with a line indicating the average rating. The ratings for all three dimensions follow a normal distribution, with the average rating being close to 6.0 for each. 
%The average ratings for aesthetics and design quality are both 6.0, while the average usability rating is 5.9. 
There are very few ratings that are lower than 4; this could be because these UI screens come from app published on the Google Play Store \cite{clay}, which went through an approval process that filtered out poorly designed apps.

We next carry out more in-depth analyses to understand the topics covered by the critiques, tasks supported by the dataset UIs, and the proportion of comments targeting UI elements, groups, and the entire screen. An analysis on the correlation between usability and aesthetics ratings can be found in Section \ref{sec:correlationanalysis} of the Appendix.

\subsection{Design Critique Topics}\label{sec:critiquetopic}
\begin{figure*}
  \centering
  \includegraphics[width=0.9\linewidth]{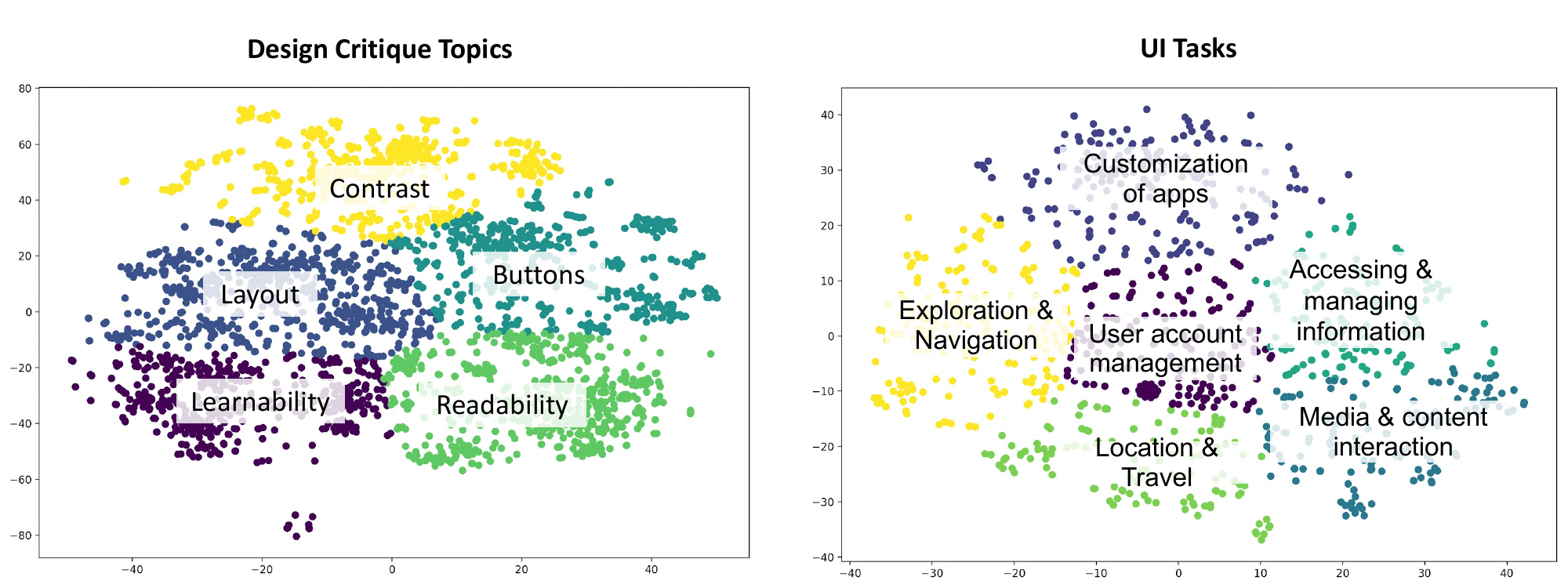}
  \caption{The results of K-means clustering by critique topic (left) and UI task category (right). Each cluster is labeled with its corresponding topic, which was determined through qualitative analysis.}
  \Description{A woman and a girl in white dresses sit in an open car.}
  \label{fig:clustering}
\end{figure*}
To qualitatively understand the dataset, we characterize the types design issues covered by the critiques. To determine the different categories of design issues, we carried out KMeans clustering of the semantically meaningful embeddings of the critique text generated by SentenceBERT \cite{reimers-2019-sentence-bert}.  We tuned the number of clusters using the Elbow method \cite{thorndike1953belongs} to 5 and reduced the dimensions to 2 using t-SNE \cite{vanDerMaaten2008} for visualization. 

Figure \ref{fig:clustering} (left) shows the results of the clustering. \changes{Two of the authors qualitatively analyzed the critiques in each cluster using grounded theory coding \cite{GlaStr67} and thematic analysis \cite{thematicanalysis} to determine the main theme for each cluster. One author coded all the design comments in each cluster, while another coded a smaller randomized sample. Each author conducted three rounds of coding to determine higher-level themes and cluster topics, and the authors met after each round to establish consistency. The codebook generated for this analysis can be found in Section \ref{sec:critiquecodebook} of the Appendix.} The two authors arrived at the following themes for each cluster:
\begin{itemize}
    \item \textbf{Layout (size: 696):} focuses on the layout of the UI screen, and includes critiques regarding positioning and alignment, the visual hierarchy, the logical grouping of elements, and simplicity of the layout.
    \item \textbf{Color Contrast (size: 655):} targets the color contrast of text, icons, buttons, etc. with the background color. 
    \item \textbf{Text Readability (size: 591):} contains critiques regarding the readability of text, based on font-size and weight.
    \item \textbf{Usability of Buttons (size: 601):} examines the usability of buttons, and includes critiques regarding the visual design of buttons for better usability, the clarity of the button's purpose, and the addition of buttons to simplify tasks (e.g. for navigation) 
    \item \textbf{Learnability (size: 601):} contains critiques regarding the clarity or intuitiveness of the purpose of icons, other UI elements, regions of the UI screen, and the entire UI screen, as well as critiques covering the clarity of text labels and other text content
\end{itemize}
The clusters in Figure \ref{fig:clustering} (left) are labeled with their corresponding themes. We also compare these clusters with the guidelines from all three sets of heuristics used for the data collection, to see if any major types of design issues are missing. One design heuristic not covered by these clusters is error prevention, and also those not applicable to the evaluation of static single screen UIs, such as heuristics feedback to the user, consistency across screens in the app, direct manipulation of UI elements, and help and documentation, which are reasonably left out of the dataset. There were two critiques in the dataset that cover visibility of system status (from Nielsen Norman's 10 Usability Heuristics), and since the sample size was so small, they were not reflected in the clusters. Furthermore, there were guidelines that were covered to a limited extent. Namely, most of the critiques related to user control and flexibility/efficiency of use involved the addition of buttons to simplify the process. The remaining heuristics were well-covered, such as metaphors, which was covered by the Learnability cluster, and all the visual design related heuristics, as four of the five clusters are relevant to visual design. 

\subsection{Tasks Supported by Dataset UIs}\label{sec:critiqueuis}
\begin{figure}
  \centering
  \changes{\includegraphics[width=\linewidth]{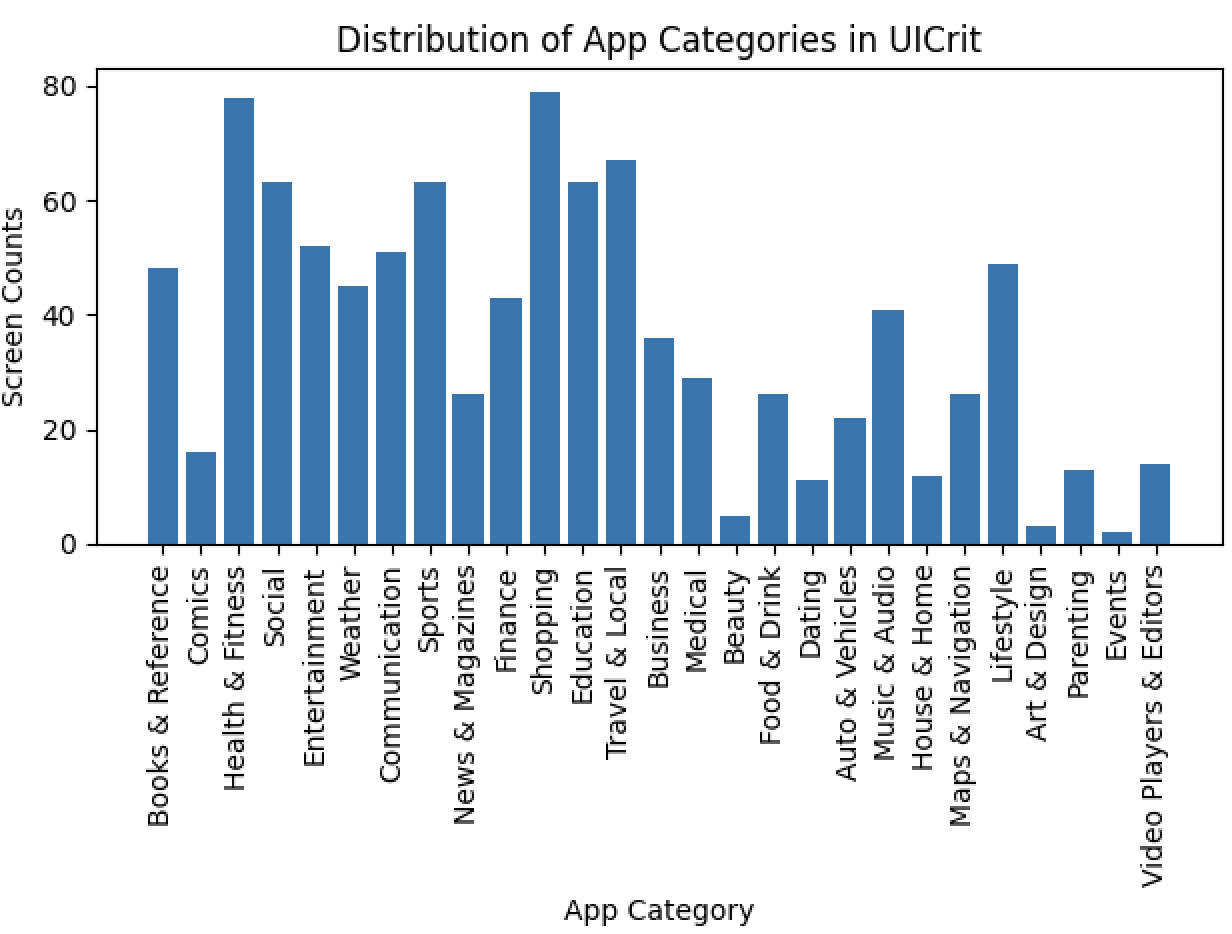}}
  \caption{\changes{Histogram showing the distribution of screens from each app category in UICrit.}}
  \Description{A woman and a girl in white dresses sit in an open car.}
  \label{fig:appcategoryhistogram}
\end{figure}
We carried out a similar analysis on the tasks annotated by workers to determine the types of tasks supported by the UIs in this dataset. This metadata is useful for selecting few-shot examples based on task similarity to the target UI, as it provides an overview of what types of tasks are represented in this dataset. 

Figure \ref{fig:clustering} (right) shows the results of this task clustering, and each cluster is labeled with its respective theme. An qualitative analysis revealed the following themes for each of the six clusters:
\begin{itemize}
    \item \textbf{Accessing and managing information (size: 144):} includes tasks like tracking health and fitness, obtaining news on the weather, sports, etc, and tracking personal expenses
    \item \textbf{User account management (size: 213):} contains tasks related to account creation and authentication 
    \item \textbf{Media and content interaction (size: 122):} includes tasks like playing music or video, and learning a new language 
    \item \textbf{Exploration and Navigation (size: 175):} contains tasks involving the exploration of apps such as browsing categories, shopping options, etc, and navigating to another page through a process (e.g. onboarding).
    \item \textbf{Customization of app settings and preferences (size 198):} includes tasks like adjusting the notification settings and setting an alarm, wallpaper, etc
    \item \textbf{Location and Travel (size: 131):} contains tasks involving navigation with maps, location tracking, and managing bookings (flights, hotels, etc) for travel 
\end{itemize}
Comparing these categories of tasks with the categories of apps from RICO \cite{Deka:2017:Rico}, a major category of tasks that is missing is communication, such as instant messaging or sending an email.

\changes{At a higher level, we could directly plot the distribution of categories for the apps that the UI screens in UICrit are from. Figure \ref{fig:appcategoryhistogram} shows a histogram of the screen counts in each app category (from RICO). This is different from the screen-level task analysis, as the screen's task may not reflect its app's category (e.g. a login screen for an Events app). According to the histogram, there are screens from all 27 apps, but the Beauty, Art \& Design, and Events categories are underrepresented.}
\subsection{Proportion of Group, Element, and Screen-level Comments}
\begin{figure}
  \centering
  \includegraphics[width=\linewidth]{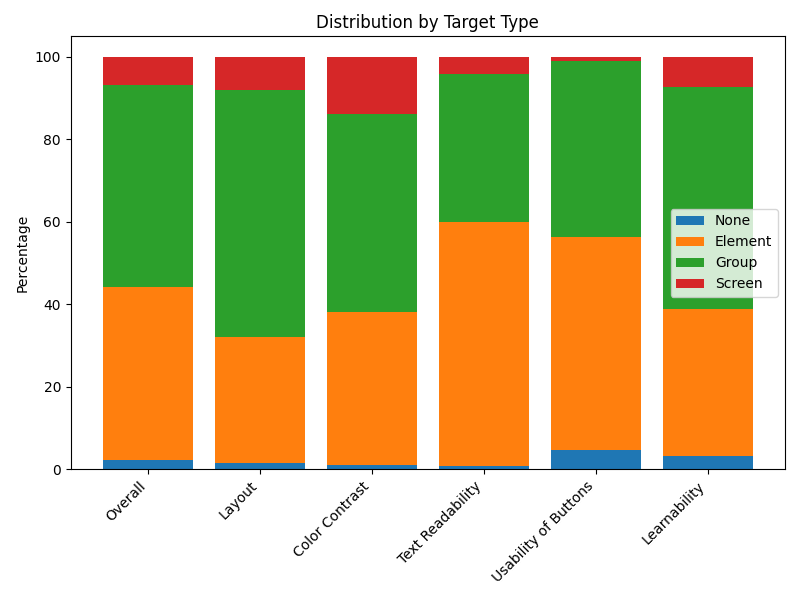}
  \caption{Stacked bar chart showing the percentage of comments targeting an individual UI element (orange), a group of UI elements (green), the entire UI screen (red), and no UI elements (blue). There is one bar for all design critiques (labeled ``Overall'') and one for each category of design critiques, labeled with the category.}
  \Description{A woman and a girl in white dresses sit in an open car.}
  \label{fig:critiquebreakdown}
\end{figure}
Another interesting way to analyze the critiques is to break them down by ones that target individual UI elements, groups of elements, and the entire screen. To compute the proportion of each type of target, we obtained location and size data for each UI element from their android view hierarchy data, which is available in the RICO dataset, to determine the number of elements in each critique's bounding box.

Figure \ref{fig:critiquebreakdown} contains a \changes{stacked bar chart} that visualizes the percentage of each type of target for all critiques (``Overall'') and critiques in each category from Section \ref{sec:critiquetopic}. For the overall distribution, there is a fairly even proportion of individual element and group annotations. Furthermore, we found a small percentage of critiques that do not target any UI elements (labeled ``none'' in the \changes{stacked bar chart}). We looked into these critiques and found that they include comments about missing elements, a location where an element should be, or commenting on a patch of the background color.

By critique topic, layout-related comments had considerably more ``group'' critiques than other types, which is expected, as these comments target layout and grouping-logic, which usually applies to groups of elements. However, there are element-level layout critiques, and they generally target individual elements with size or positioning issues. Contrast-related comments generally match the overall distribution of target types, as they cover contrast issues of individual elements and groups with the background. Contrast critiques have the highest percentage of screen-level comments out of all categories, which could be due to comments regarding poor background color choice for the screen. There were considerably more element-level comments regarding text-readability, which could be because they usually target the readability of individual elements, though they could sometimes apply to groups of text (e.g. items in a menu), which take up a third of the comments. We expected button usability critiques to mostly target elements, but the significant portion of group-level comments could come from comments targeting groups of buttons, which are common. Button usability comments also have the highest percentage of critiques targeting no elements, and they generally come from comments regarding the addition of a new button to improve usability. Finally, learnability-related comments also follow a similar distribution of target types as the overall distribution, which is expected as they target the intuitiveness or elements, groups, and the entire screen. There are a few deviations from expectation, like a small fraction of text-readability comments that target no elements, which could be attributed to imperfect clustering. 

\subsection{App-level Analyses}
\subsubsection{Correlation Between UI Screen Ratings and App Ratings}
\changes{We computed the Pearson correlation coefficient between the screen-level ratings for aesthetics, usability, and overall design quality from UICrit and the app ratings on the Google Play Store (taken from RICO \cite{Deka:2017:Rico}), for the apps the screens are from. For apps with multiple screens in the dataset, we averaged the ratings for those screens.} 

\changes{We obtained the following correlation coefficients: 0.007 (Aesthetics), 0.022 (Usability), and 0.023 (Overall Design Quality). Each type of rating had a weak positive correlation with the app rating. While this result deviates from expectation, it could be due to fact that the UICrit evaluation ratings were based on one or a few screens from the app, which may not represent the entire app’s design. Additionally, app ratings are influenced by many other factors such as app latency, spam/ads, cost, and customer service \cite{NAKAMURA2022111462}.}

\subsubsection{Consistency of Screen Ratings from the Same App}
\begin{table}[t]
\centering
\begin{tabular}{ccc}
\toprule
\textbf{Rating Type} & \textbf{Avg. App-level SD} & \textbf{Dataset-level SD}\\ \midrule
\changes{Aesthetics} & \changes{0.65} & \changes{1.17} \\ \hline
\changes{Usability} & \changes{0.64} & \changes{1.25}\\ \hline
\changes{Design Quality} & \changes{0.60} & \changes{1.13} \\ \hline
\bottomrule
\end{tabular}
\caption{\changes{The standard deviation of ratings for screens with the same app (``Avg. App-level SD'') and for screens in the entire dataset (``Dataset-level SD'') for each rating type.}}
\label{tab:ratingconsistency}
\end{table}
\changes{We also measured the consistency of the UICrit ratings for screens from the same app. We found 95 apps with at least two screens in the dataset, with a total of 208 screens. To measure consistency, we computed the standard deviation of the ratings for UI screens within the same app and compared it with the standard deviation of the ratings for the entire dataset. The average standard deviation across these 95 apps for each rating type, along with the standard deviation of the entire dataset for that rating type, can be found in Table \ref{tab:ratingconsistency}. The table shows that the average app-level standard deviation is around half that of the standard deviation for the entire dataset, for all rating types. This indicates some consistency among screen ratings within the same app.}

\section{Application of Dataset for Modeling UI Feedback Generation}
\begin{figure*}
  \centering
  \includegraphics[width=\linewidth]{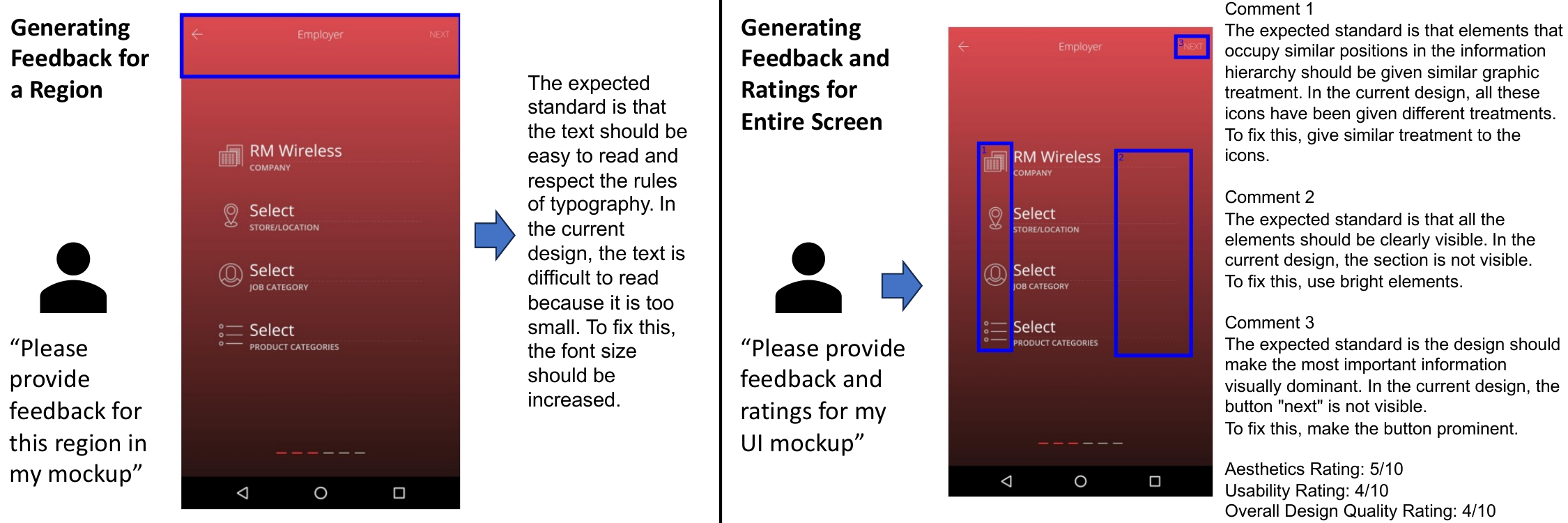}
  \caption{Illustration of the input and output for each modeling task. The left part of the figure illustrates the generation of comments for a specific region in the screenshot (marked), and the right part shows the generation of critiques, corresponding bounding boxes, and design quality ratings for the entire UI screen. \changes{These are realistic outputs from our best performing few-shot setup for each task, which sometimes contain hallucinations}}
  \Description{A woman and a girl in white dresses sit in an open car.}
  \label{fig:taskillustration}
\end{figure*}
The data available in UICrit can be applied to help automate two distinct feedback generation tasks that are potentially useful to designers: 1) generating feedback for a specific region in the UI screen that the designer is concerned with (Figure \ref{fig:taskillustration}, left) and 2) producing feedback for the entire UI screen. The latter includes automatically marking the target region (bounding box) of each critique, and also generating design quality scores (Figure \ref{fig:taskillustration}, right). We utilized this dataset to experiment with various few-shot training and visual prompting methods to tune Gemini for each task. We then ran a user study, where a different group of design experts (who did not participate in the critique data collection) compared the feedback generated by Gemini with our few shot and visual prompts against the feedback from general-purpose (zero-shot) Gemini and human annotated feedback from the dataset. 
\subsection{Generating Design Comments for a Region of Interest}
This task entails generating design feedback specifically for a region in the UI (Figure \ref{fig:taskillustration}), and addresses scenarios where a designer would want to obtain targeted feedback on a particular portion of the screen design.
\subsubsection{Prompt Design}
The prompt design includes the UI screenshot overlaid with a bounding box around the region of interest (ROI), and the instructions to provide UI feedback for the marked region and how to properly format the feedback, following \cite{sadler1989formative}. We also provide the three sets guidelines that were used during the data collection and ask the LLM to find guideline violations to supplement its feedback. 
\subsubsection{Few Shot Method}\label{sec:samplingtypes}
We tried three techniques to select few shot samples from UICrit: random sampling, sampling by visual similarity, and sampling by semantic similarity. Random sampling entails selecting a random UI and then a random bounding box from one of its critiques as the region of interest, with the corresponding critique as the ground truth output. Selecting a critique's bounding box ensures that the ROI corresponds to a meaningful region or element in the screenshot. %While random sampling would likely select few shot examples that are less similar that the two targeted approaches, random sampling may select more diverse few shot samples, which could potentially lead to a more diverse set of comments. 
Sampling by visual similarity involves selecting bounding box patches (cropped from the UI screenshot) that are the most visually similar to the input UI's region of interest. The intuition behind providing visually similar few shot examples is that UI regions that are visually similar may have similar critiques (especially regarding the visual design), which increases the chance of the LLM providing relevant and accurate feedback. We used root mean square difference to compute the visual similarity 
\cite{rmse} between bounding box patches from the dataset and the ROI's patch. We then select the UIs, bounding boxes, and corresponding critiques from UICrit containing the most visually similar patches as few-shot examples. Sampling by semantic similarity entails applying CLIP \cite{radford2021learning} to embed the patch image in a shared image and text space. CLIP is optimized for finding the closest text description for an image, which means this CLIP embedding would capture the semantic details of the input patch, for instance, if it contains a login button. The intuition for semantic sampling is that it would likely select patches that are semantically similar to the target patch that would provide more relevant critiques.
\subsubsection{Few Shot Results}\label{sec:outputeval}
\begin{table}[t]
\centering
\begin{tabular}{p{2.5cm}p{2.5cm}p{2.5cm}}
\toprule
\textbf{Few Shot Setup} & \textbf{Avg. Comment \newline Quality Score} & \textbf{Total Number of Comments}\\ \midrule
\changes{8-shot, \newline Visual Similarity} & \changes{0.44} & 11 \\ \hline
\changes{8-shot, \newline Semantic Similarity} & \changes{0.42} & 10\\ \hline
\changes{4-shot, \newline Visual Similarity} & \changes{0.36} & 20 \\ \hline
\changes{4-shot, \newline Semantic Similarity} & \changes{0.34} & 21\\ \hline
\changes{8-shot, Random} & \changes{0.31} & 9\\ \hline
\bottomrule
\end{tabular}
\caption{Results of the few-shot experiments for the top 5 performing set-ups for the task of generating design comments for a region of interest in the screenshot. The table shows the average quality score per comment \changes{(from all 3 annotators)}, as well as the total number of comments generated for the 6 UIs for each setup.}
\label{tab:patchfewshot}
\end{table}
To evaluate each few shot method, we applied Gemini to generate UI feedback for bounding box patches corresponding to critiques using 2, 4 and 8 shots. Eight was the maximum number of shots for the prompt to consistently not exceed the context window limits of Gemini Pro Vision. We then devised a scoring method to evaluate the output, where a valid critique was assigned 1 point, a partially valid critique was assigned 0.5 points, and in invalid critique was assigned 0 points. A critique is considered valid if it is both accurate and helpful, following the rating criteria described in \cite{duan2024}. \changes{We recruited three annotators to manually score these design comments. Two of the annotators were authors with prior design experience, and the third annotator was P1 from Table \ref{tab:expertdetails}} 

Table \ref{tab:patchfewshot} shows the scores for the top 5 performing few-shot configurations, applied to 6 distinct UI screens. \changes{The scores are normalized by the total number of comments (shown in a separate column) and averaged across all three annotators, allowing for comparison across different few-shot methods.} \changes{The three annotators had a Fleiss' Kappa \cite{fleiss1971mns} inter-rater agreement score of 0.30, which indicates fair agreement.} \changes{Visual sampling} with 8 shots had the best performance, and semantic and visual sampling outperformed random sampling, even with fewer shots, which illustrates the effectiveness of these targeted sampling methods. One interesting observation to note from the table is that, when there are fewer shots, Gemini generates more comments for each patch, \changes{but they are of lower quality (on average)}. This case is more apparent for 0-shot, which had 45 comments total and a score of 0.24. \changes{The model likely learns to give comments of higher quality for the target patch from the few shot examples.} 
\subsection{Generating Design Comments for Entire UI Screen with Score Prediction}
Since modeling UI screen comments generation and UI screen comments and rating generation are the same, other than a slight modification in the prompt, we focus on the second task. Automated critique and design rating generation is useful in cases when a designer would like feedback for the entire UI screen mockup, and automatic design prediction could be used to quantitatively compare different UI designs or to see if a revision improved the design. 
\subsubsection{Prompt Design}
Similar to the prompt for the previous task, the prompt for this task includes the UI screenshot (without any bounding boxes), the three sets of guidelines, and instructions on properly formatting the critique. Unlike the previous task, this prompt asks the LLM to provide critiques for the entire UI screen and to provide bounding boxes for each critique highlighting relevant regions in the screenshot.
\subsubsection{Few Shot Method}
We tried four different methods to select few shot samples from the dataset: random sampling, sampling based on visual similarity, sampling based on task similarity, and sampling based on visual and task similarity. Random and visual similarity sampling follow the same procedure as described in Section  \ref{sec:samplingtypes}, except visual similarity comparison is now carried out on screenshots, and the UI screen and its entire set of critiques are selected, instead of those corresponding to a single region. Task similarity sampling selects UI screens with the most similar tasks, and the intuition is that UIs with similar tasks would likely have similar usability requirements and the critiques from few shot examples with similar tasks would be informative regarding the usability requirements of the input UI. Task similarity is computed by using SentenceBERT to embed the task description and then computing cosine similarity on the embeddings. Sampling based on task and visual similarity combines the two sampling methods into a more comprehensive sampling approach. We used CLIP to generate a joint embedding of the UI screenshot and task description and then apply cosine similarity on the embeddings to determine similarity. We did not use this joint task and visual similarity sampling method for generating design comments for an ROI because the task for the UI screen may not be semantically relevant to the target region in the UI screen (i.e. if it only contains an icon), which could introduce noise. 
\subsubsection{Few Shot Results}\label{sec:fewshotui}
\begin{table}[t]
\small
\centering
\begin{tabular}{p{2.5cm}p{2cm}p{1.5cm}p{1.5cm}}
\toprule
\textbf{Few Shot Setup} & \textbf{Avg. Comment \newline Quality Score (without bbox)} & \textbf{Total No. of Comments} & \textbf{Avg. Rating Accuracy Score}\\ \midrule
8-shot, Visual and Task Similarity & \changes{0.58} & 14 & 0.53\\ \hline
\changes{4-shot, Visual and Task Similarity} & \changes{0.48} & 10 & 0.22\\ \hline
\changes{8-shot, Task Similarity} & \changes{0.42} & 8 & 0.31\\ \hline
\changes{4-shot, Task Similarity} & \changes{0.38} & 12 & 0.17\\ \hline
\changes{8-shot, Visual \newline Similarity} & \changes{0.38} & 10 & 0.17\\ \hline
\bottomrule
\end{tabular}
\caption{Results of the few-shot experiments for the top 5 performing set-ups for the task of generating comments, comment bounding boxes, and design quality ratings for the entire UI screen. The table shows the average quality score per comment (without considering bounding box accuracy), the total number of comments, and the average rating accuracy score for the 6 UIs for each setup. We did not evaluate bounding box accuracy at this stage, as they will be tuned at a later step with visual prompting.}
\label{tab:screenfewshot}
\end{table}
We reused the evaluation method from \ref{sec:outputeval} \changes{and the same three annotators} to determine the best few-shot scoring method. For this evaluation, we also generate scores for the predicted aesthetics, usability, and design quality ratings based on the ground truth ratings from the dataset. We assign a score of 1 if the ratings match, a score of 0.5 if the predicted rating deviates by 1 point from the ground truth rating, and a score of 0 otherwise. We do not evaluate predicted bounding box accuracy at this stage, as we tune bounding boxes in a later step.

Table \ref{tab:screenfewshot} shows the top 5 performing few-shot configurations (based on normalized comment quality score). \changes{The three annotators had a Fleiss' Kappa \cite{fleiss1971mns} inter-rater agreement score of 0.31, indicating fair agreement.} \changes{The joint task and visual similarity sampling method had the best performance with both numbers of shots, followed by task similarity (both shot counts), and then 8-shot visual similarity, which tied with 4-shot task similarity sampling.} The fact that joint visual and task similarity sampling has the best performance out of all sampling methods aligns with expectation, because this method likely samples UI screens that are the most similar overall to the target UI. Furthermore, as with the ROI comment generation task, the targeted few shot sampling methods all outperformed random sampling, where 8-shot did not even make the top 5. Finally, the joint task and visual similarity sampling method greatly outperformed the other methods (with the same number of shots) in predicting the numerical design quality ratings based the normalized rating scores in Table \ref{tab:screenfewshot}.

\subsubsection{Visual Prompting Method}
Since LLMs have poor object localization \cite{dorkenwald2024pin}, we tried various ``visual prompting'' methods \cite{liu2024moka} to improve the accuracy of the corresponding bounding boxes generated for each design critique. Our visual prompting approaches include adding coordinates on the edge of the UI screenshot to assist in the specification of bounding boxes coordinates, overlaying a grid over the screenshot, and overlaying a patch grid over the screenshot, where the LLM would just need to return the numbers of the patches corresponding to the bounding box, which was done by \cite{liu2024moka}. Figure \ref{fig:visualprompting} in the Appendix illustrates each visual prompting method.
\subsubsection{Visual Prompting Results}
\begin{table}[t]
\centering
\begin{tabular}{p{4cm}p{3.5cm}}
\toprule
\textbf{Visual Prompt Method} & \textbf{Avg. Bounding Box IoU}\\ \midrule
Screenshot Only & 0.004\\ \hline
Screenshot with Grid & 1.1e-5\\ \hline
Screenshot with Patches & 0.222\\ \hline
Screenshot with Coordinates & 0.186\\ \hline
\bottomrule
\end{tabular}
\caption{Results of the different visual prompting techniques to localize comment bounding boxes. The table shows the average intersection over union score per bounding box for each method.}
\label{tab:visualprompt}
\end{table}
\begin{figure*}
  \centering
  \includegraphics[width=0.9\linewidth]{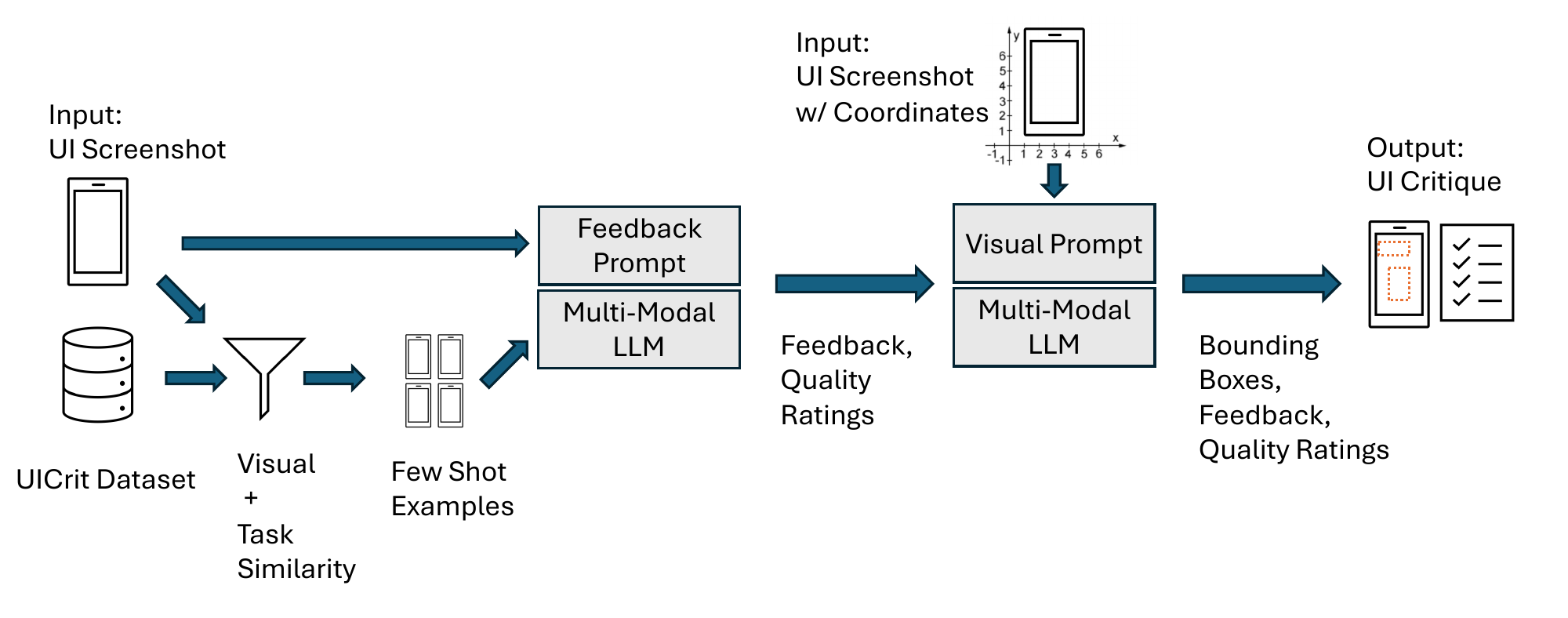}
  \caption{The diagram illustrates the optimal prompt chain setup, which consists of few-shot prompting to obtain UI comments and design quality scores, followed by a prompt that uses visual prompting methods to localize comment bounding boxes in the input UI screenshot.}
  \Description{A woman and a girl in white dresses sit in an open car.}
  \label{fig:promptchain}
\end{figure*}
We first tried combining critique generation and bounding box detection with visual prompting into a single LLM call, but found that while the bounding boxes improved in accuracy with visual prompting, the design critiques decreased in quality. This is probably due to task overload on the LLM. We split critique generation and bounding box detection into two separate calls to preserve the critique quality, while improving bounding box accuracy. The second prompt takes the critique output from the previous LLM call and queries the LLM with the critiques, screenshot (with various visual prompting methods), and instructions to output the corresponding bounding box in the screenshot for each critique. This chain of prompts, one for each major task, is illustrated in Figure \ref{fig:promptchain} and follows the prompt design from \cite{duan2024}.

To evaluate bounding box quality, we compute the intersection of union (IoU) between the output bounding box and the ground truth bounding box for the comment (manually determined by the authors). For a fair comparison, we ran the bounding box detection prompts on the generated design critiques from the best few-shot configuration (8-shots with joint task and visual similarity sampling) from Section \ref{sec:fewshotui}. Table \ref{tab:visualprompt} illustrates the average bounding box IoU from various visual prompting techniques. The ``Screenshot with Grid'' visual prompting technique had the worst performance, even lower than the output without any visual prompting (``Screenshot Only''). This could be attributed to the fact that the gridlines overlaid on the screenshot introduced visual noise. While the patch grid is also overlaid, outputting the patch numbers corresponding to bounding boxes is easier than estimating coordinates, which may have offset the visual noise. Despite the fact that using coordinates had slightly worse performance than the patches condition, we used its bounding boxes for the user study in Section \ref{sec:userstudy} due to the fact that the Patches condition sometimes returned irregular (i.e. non-rectangular) ``bounding boxes'', and that the bounding boxes are not as precise as those from the Coordinates condition. Figure \ref{fig:promptchain} illustrates the complete prompt chain for this setup.
\subsection{Validation of Performance Improvement}\label{sec:userstudy}
To validate that our few-shot prompt design with usage of UICrit actually results in improved performance, we ran a user study on the design critiques generated from Task 2 (generating UI comments and corresponding bounding boxes for the entire UI), as it is a broader use case compared to generating targeted feedback for a specified region in the UI (Task 1). We also leave out design quality scores from the evaluation, as we were able to compare it with the ground truth design quality ratings from the dataset. 
\subsubsection{Method}
\begin{table}[t]
\centering
\begin{tabular}{|c|c|c|}
\toprule
\textbf{Participant} & \textbf{Design Expertise} & \textbf{Yrs. of Exp.}\\ \midrule
\changes{P1} & \changes{Interaction} & \changes{9} \\ \hline
\changes{P2} & \changes{UX, Product, Visual, Graphic} & \changes{30}\\ \hline
\changes{P3} & \changes{UI/UX, Interaction, Visual, Graphic} & \changes{9} \\ \hline
\changes{P4} & \changes{UI} & \changes{8}\\ \hline
\changes{P5} & \changes{UI} & \changes{10}\\ \hline
\changes{P6} & \changes{Interaction} & \changes{5}\\ \hline
\bottomrule
\end{tabular}
\caption{\changes{The areas of design expertise and number of years of professional design experience for each of the 6 design experts who participated in the validation user study.}}
\label{tab:expertdetails}
\end{table}
We use the best performing configuration from Task 2: 8-shot with joint task and visual-similarity sampling for critique generation, and using visual prompting with coordinates to generate corresponding bounding boxes, and compare it with two baselines: zero-shot prompting for critique generation, followed by another LLM call to obtain corresponding bounding boxes without visual prompting and human generated feedback from the dataset. We make two LLM calls for the first baseline, following that finding from \cite{duan2024} that separating LLM calls for major tasks leads to better UI design feedback. 
We recruited 6 design experts from our institution for this study. \changes{Table \ref{tab:expertdetails} details the areas of design expertise and years of professional design experience for each participant.} During the study, participants evaluated the design feedback and bounding boxes from all three cases for one UI at a time, for a total of 6 UIs. For each UI, they first scored each UI comment individually, following the scoring method detailed in \ref{sec:outputeval}. For the second part, they evaluated the set of comments from each case as a whole, and ranked the sets of comments based on their overall quality and comprehensiveness. Participants also provided explanations for their rankings. \changes{We provided a form for participants to record their scores, rankings, and ranking explanations. Before starting the study's tasks, we held a brief meeting with each participant to explain instructions and obtain consent.}
\subsubsection{Results}
\begin{figure}
  \centering
  \includegraphics[width=\linewidth]{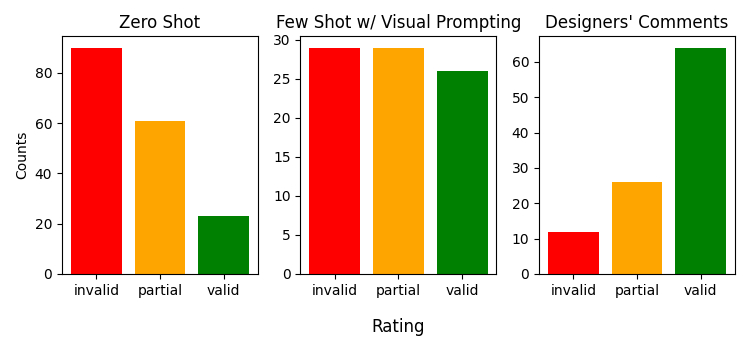}
  \caption{Histograms showing the distribution of expert ratings on the quality of design critiques from the zero shot method, the few-shot with visual prompting method, and from designers' (taken from the dataset).}
  \Description{The rating distribution of critique quality}
  \label{fig:userstudyratingdistribution}
\end{figure}
\begin{table}[t]
\centering
\begin{tabular}{p{3cm}p{2cm}p{1.5cm}p{1cm}}
\toprule
\textbf{Condition} & \textbf{Avg. \newline Comment Quality Score (with bbox)} & \textbf{Total Number of \newline Comments} & \textbf{Avg. Rank}\\ \midrule
Zero-shot & 0.31 & 29 & 2.6\\ \hline
8-shot, Visual and Task Similarity, Coordinates Visual Prompting & 0.48 & 14 & 2.1 \\ \hline
Human & 0.75 & 17 & 1.3\\ \hline
\bottomrule
\end{tabular}
\caption{Quantitative results from the user study comparing the comments generated from the best few-shot with visual prompting configuration against those from zero-shot prompting and human designers. The table includes the average comment quality score (with bounding box accuracy included in the evaluation), the total number of comments, and the average preference ranking (out of the 3 different conditions). Note that the average comment quality score is lower than those from Table \ref{tab:screenfewshot} because bounding box accuracy is also included in the evaluation, and incorrect bounding boxes would lower the score.}
\label{tab:userstudyresults}
\end{table}
\changes{To compare comment quality across the three cases, we again normalized the total comment quality score by the number of comments generated by each condition and then averaged this score across the 6 participants.} Table \ref{tab:userstudyresults} shows the average comment quality score for our setup (``Few-shot with Visual Prompting'') and the two baselines, along with the total number of comments from each condition. These results align with expectation, where our setup outperformed the zero-shot baseline by 0.17 ($p = 5\mathrm{e}{-4}$) in average quality score, which corresponds to a 55 percent increase. As expected, the critiques from humans had the highest average quality score. However, the quality score is less than 1, which is likely due to the fact that design evaluation is inherently subjective, leading to disagreement regarding design issues present in the UI. In fact, we computed the Fleiss Kappa inter-rater reliability score \cite{fleiss1971mns} for the critique ratings across participants and obtained a value of 0.29, which implies fair agreement. We also visualize the rating distributions for each condition in Figure \ref{fig:userstudyratingdistribution}, which shows that the Zero Shot condition had the highest fraction of invalid ratings and a very low fraction of valid ratings. Our setup had a fairly uniform distribution, with a slightly smaller proportion of valid ratings, and the majority of human critiques had valid ratings. This plot and table \ref{tab:userstudyresults} also shows that zero-shot has considerably more comments than the other two conditions, which could be caused by Gemini's hallucinations, as indicated by the high number of invalid ratings. \changes{This implies that our few-shot training method probably reduced hallucinations, compared to the zero-shot setup.}

\changes{We also computed the average ranking number for the set of comments generated by each condition across all UIs and participants. The ranking number for the best comment set would be 1, the second-best set would be 2, and the worst set would be 3. Therefore, lower ranking numbers indicate a better set of comments (i.e., a higher ranking in the group). Table \ref{tab:userstudyresults} shows the average rank number for all three conditions.} Our setup had a higher average ranking than the zero-shot condition, and the set of human comments was generally ranked the highest. The ranking data had an agreement score of 0.55, which indicates moderate agreement. The participants preferred the set of critiques generated by our setup over the zero-shot set of critiques 67 percent of the time. Their reasons for preferring our setup's comment set (besides higher accuracy), include greater comment specificity to the design (P1, P3), more actionable feedback (P1), and the fact that zero shot comment set made too many assumptions (P3, P5). The times when the zero shot comment set ranked higher than our setup's was justified by reasons like the feedback was easier to understand (P3, P4) and the set of comments was more thorough (P4). However, the participants usually prefer the set of human critiques out of all three sets, which was ranked higher than the critique set from our setup 81 percent of the time. Reasons for preferring human critiques include higher relevance to the UI (P3, P5), more detailed guidance (P1, P3, P5), and providing the best design rationale (P4). We obtained these reasons through qualitatively analyzing the ranking explanations.

\section{Discussion}
We discuss insights from our few-shot experiments and the user study on the utility of this dataset, potential broader applications of this dataset, \changes{and the scalability of our data collection method}.
\subsection{Utility of the Dataset}
Results from the few-shot experiments show that a targeted few shot sampling approach performed better than random sampling. This dataset probably contributed to the strong performance of these targeted few shot sampling methods because it provided a large set of UIs to sample from, which likely resulted in few shot examples that are quite similar visually, semantically, or both to the input UI screen, which provided more informative critiques for its assessment. This implies that this dataset, combined with few shot sampling methods that fully utilize it, enables the generation of better UI feedback compared to methods that do not use this dataset.

The improvement in generated comment quality was confirmed via a user study with design experts, who scored the comments generated by the best performing targeted few shot sampling and visual prompting setup 55 percent higher than the comments generated by zero-shot prompting, which corresponds to the condition of not utilizing dataset. Furthermore, the design experts also more often preferred the set of comments generated for a UI by the few-shot and visual prompting condition over the comments generated by the zero-shot condition. While this dataset enables better automated design feedback, the design feedback is still far from the quality of the design critiques from humans, which was rated and ranked considerably higher than the few-shot and visual prompting condition. While there is still a significant performance gap between this 8-shot with visual prompting technique and experienced human designers, fine-tuning an LLM on all 938 UI examples in this dataset may potentially lessen this gap.  
\subsection{Potential Applications of Dataset}
In addition to direct applications for few-shot training and fine-tuning LLMs and other models, this dataset has potential broader applications in the field of computational UI design.
\subsubsection{Tool-agnostic UI Evaluation}
This dataset could be applied to fine-tune a multi-modal LLM to automatically generate design critiques and ratings given only the UI screenshot. This has strong implications for its flexible integration into design tools. Only the image of the UI mockup would be needed for the fine-tuned model to generate feedback, and this universal representation should be available in all design tools, which may have different internal representations of their mock-up, such as Figma's JSON representation based on the layers configuration. This implies that the fine-tuned model can be used to evaluate the mock-up of any design tool, independent of their internal mockup representation, and can be easily integrated into any design tool that support the addition of add-ons, such as Figma, which allows the creation of plugins. Other implications include the straightforward integration of this fine-tuned model as an automated mockup evaluation feature in the development of future design tools, and the use of this model to evaluate any arbitrary UI screenshot found on the web or a UI dataset. Furthermore, since the evaluation is automated, this model enables UI evaluation at a large scale.
\subsubsection{Reward Signals for Improving UI Generators} 
Recently, significant progress has been made on UI generation using diffusion models\cite{cheng2023play, inoue2023layoutdm} and LLMs\cite{lin2024layoutprompter}. However, the generated UIs often fall short on following detailed design principles, capturing latest style trends, or avoiding artifacts like misaligned elements. To enhance the generation quality, we can fine-tune existing models with our dataset. Specifically, recent studies\cite{lee2023rlaif, xu2023reasons} have shown that natural language feedback generated by LLMs, such as the design comments in our case, can serve as reward signals for fine-tuning. Moreover, these methods are general approaches to improve model performance, making this dataset valuable regardless of the type of UI generators.
\subsection{Scalability of Data Collection System}
\changes{The data collection system described in Section \ref{sec:datacollection} is designed to be highly scalable and flexible, capable of accommodating any set of UI screenshots and design guidelines, given a sufficient group of human annotators with prior design experience. This flexibility implies that this data collection system could be applied to collect relevant critiques for new UI trends, given updated screenshots and guidelines, such as those from the release of a new, up-to-date UI dataset and current design guidelines. Running this annotation system regularly to integrate new UI trends would ensure that the dataset remains continually updated and relevant.}

\section{Limitations and Future Work}
\changes{We discuss some limitations of our work. Regarding the dataset, we had only seven participants recording design comments, and this small annotator pool may restrict the diversity of critiques in the dataset. This limitation is further evidenced by the underrepresented design issues we found in Section \ref{sec:critiquetopic}. Another dataset-related limitation is that it contains critiques relevant to only single UI screens, which leaves out feedback applicable to the entire app or UX-related feedback for task flows, such as how the app handles user errors.} Furthermore, to illustrate light-weight use cases for the dataset, we only applied the dataset for few-shot prompting and provided at most 8 UI samples from the dataset due to context window limitations. \changes{This light-weight few shot pipeline still sometimes hallucinates}, and we may miss out on a potentially larger performance gain from fine-tuning Gemini on the entire dataset. Finally, the critiques generated by our few shot and visual prompting setup were only evaluated by participants for validity. Participants in our study did not implement any of the critiques, which prevented assessment the generated critiques' helpfulness in practice and effect on design outcome.

Opportunities for future work include using this dataset to finetune multi-modal LLMs, such as Gemini Pro Vision, and evaluating the resulting performance. In addition, various input modalities could be explored, such as screenshot-only input and supplementing the screenshot input with an XML representation, and comparing their performance. \changes{Furthermore, a study could be conducted to evaluate the performance of our few-shot techniques across different multimodal LLMs (e.g., GPT-4V \cite{openai2024gpt4technicalreport}). This would help determine if the performance gains we observed for Gemini generalize across various language models.} The data collection could be expanded to include more UI screens to ensure representation of all major UI tasks, and new designers could be added to the dataset annotation pool, which may introduce more variety in the types of design issues covered in the dataset and alleviate missing or underrepresented categories of design issues. The data collection system could be extended so that workers evaluate a series of UI screens corresponding to task traces and provide UX-related critiques and ratings on the design of the task flow. There are several datasets containing UI task traces, such as \cite{rawles2023android}, that could supply task traces for this annotation. Finally, a model trained on this dataset (with only the UI screenshot as input) could be integrated as a plugin in existing design tools, such as a Figma, Sketch, and Adobe XD, which could then be used to carry out a study evaluating the generated critiques helpfulness in design practice and effect on design outcome. Given the tool-agnostic nature of the fine-tuned model, this study could be carried out on multiple design tools to see if results vary by tool.
\section{Conclusion}
We collected a dataset of design critiques, corresponding bounding boxes, and design quality ratings from experienced designers for a set of 983 distinct UIs, through a carefully constructed protocol. We then analyzed the dataset to characterize the types of design issues covered by the critiques, tasks supported by UIs in the dataset, and other informative features. We then applied this dataset to automate UI feedback generation, with a novel prompt design that includes targeted few-shot sampling from the dataset and visual prompting to determine corresponding bounding boxes for each critique. We verified that this method generates higher quality feedback compared to zero-shot prompting via a user study with design experts, which confirms that this dataset's utility in improving automated design feedback. In addition to this demonstrated contribution, UICrit has numerous potential applications, such as training a tool-agnostic model that could be integrated into any design tool, and the dataset's design critiques and quality ratings could be used to train a reward model for generative UI models. 

\bibliographystyle{ACM-Reference-Format}
\bibliography{sample-base}

%%
%% If your work has an appendix, this is the place to put it.
\appendix

\section{Correlation between Usability and Aesthetics Ratings}\label{sec:correlationanalysis}
To see if there is a correlation between the quality of usability and aesthetics for UIs in the dataset, we computed the Pearson correlation coefficient between the aesthetics and usability ratings. We obtained a correlation coefficient of 0.875, which indicates a very high positive correlation. This strong positive correlation could be attributed to the fact that poor aesthetics would lead to poor usability of the UI, and that designers who invest significant effort into aesthetics or usability would likely also put a lot of effort in the other, aiming to achieve an overall high quality UI design.

\section{Visual Prompting Input}
\begin{figure*}
  \centering
  \includegraphics[width=\linewidth]{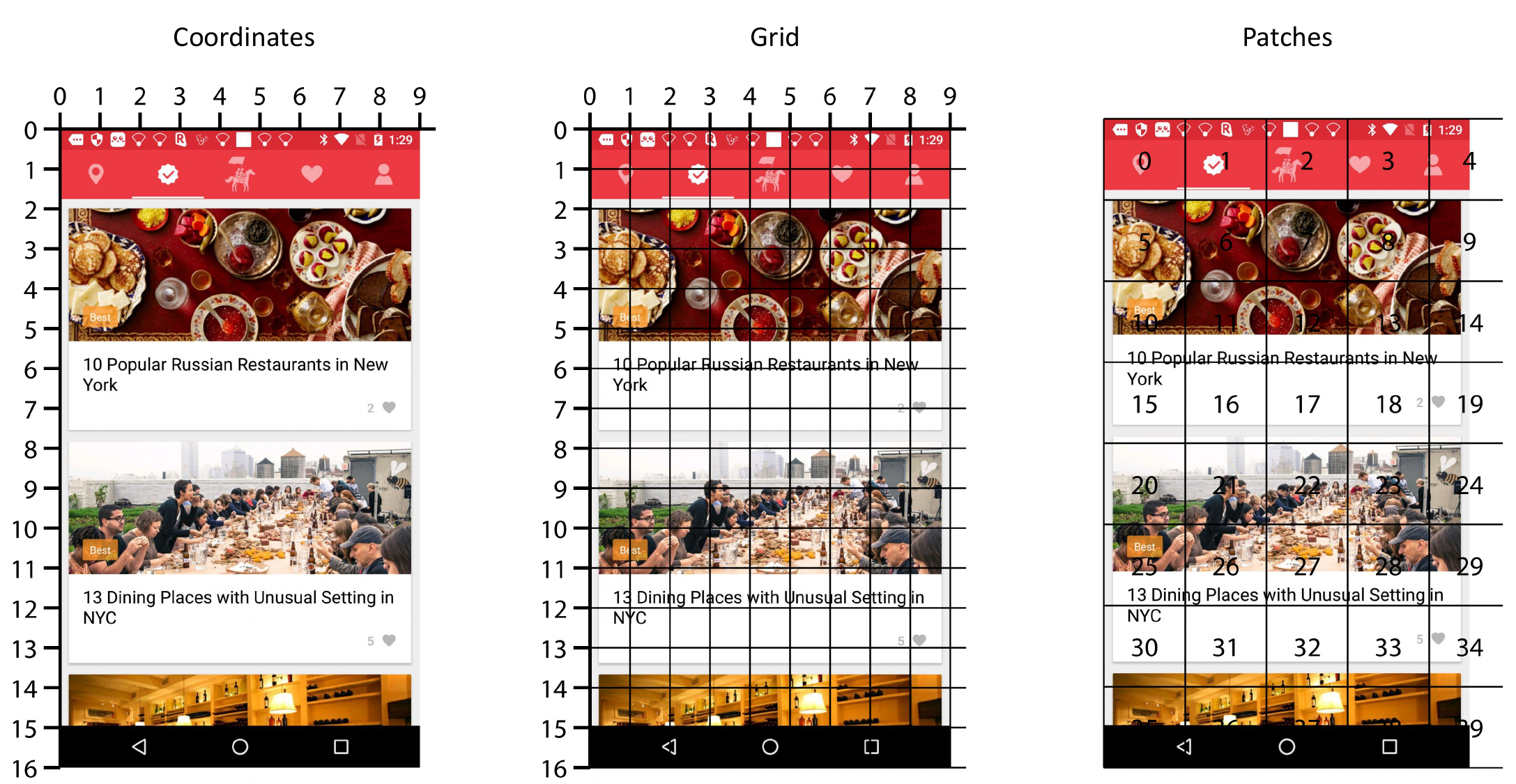}
  \caption{Illustration of the three visual prompting techniques we used to improve critique bounding box accuracy. The ``Coordinates'' visual prompting method entails adding a coordinate axis along the edges of the screenshot to help the LLM estimate the bounding box coordinates, the ``Grid'' method involves overlaying a grid over the screenshot to further help with bounding box location estimation, and the ``Patches'' method entails overlaying a grid of numbered patches over the screenshot. For the ``Patches'' method, the LLM just needs to return the numbers of patches corresponding to the bounding box.}
  \Description{A woman and a girl in white dresses sit in an open car.}
  \label{fig:visualprompting}
\end{figure*}

\changes{Figure \ref{fig:visualprompting} illustrates the three visual prompting techniques used to improve the critique bounding box accuracy.}

\section{Design Critique Topics Codebook}\label{sec:critiquecodebook}
\changes{This section contains the codebook for the qualitative analysis described in Section \ref{sec:critiquetopic}. The codebook contains the codes agreed upon at the end of each round of coding for the first two rounds, as the cluster topic and definition for the final round are discussed in Section \ref{sec:critiquetopic}. Each code is followed by its definition.}
\subsection{Layout}
\subsubsection{Round 1}
\changes{\begin{itemize}
    \item \textbf{Cluttered Layout:} Overcrowded design elements causing visual confusion.
    \item \textbf{Alignment Issues:} Misaligned text, images, or other elements.
    \item \textbf{Visual Hierarchy Problems:} Lack of differentiation in size, color, or spacing to indicate importance.
    \item \textbf{Margins and Spacing Issues:} Inconsistent or insufficient spacing between elements.
    \item \textbf{Redundant or Unnecessary Elements:} Elements that do not contribute to the design's purpose or user experience.
    \item \textbf{Poor Text Justification and Formatting:} Poor text alignment and justification
    \item \textbf{Lack of Visual Emphasis on Interactive Elements:} Interactive elements that do not stand out or are not clearly marked.
    \item \textbf{Visual Disorganization:} Lack of a clear structure or logical arrangement of elements.
    \item \textbf{Redundant Information:} Excessive or repetitive information overwhelming the user.
    \item \textbf{Ineffective Use of White Space:} Poor management of white space leading to disjointed or cluttered appearance.
\end{itemize}}
\subsubsection{Round 2}
\changes{\begin{itemize}
    \item \textbf{Position and Alignment:} Proper placement and alignment of design elements
    \item \textbf{Visual Hierarchy:} Strategic use of size, color, spacing, and other visual features to establish a clear hierarchy of information.
    \item \textbf{Logical Grouping of Elements: The org}anization and logical arrangement of design elements into coherent/related groups.
    \item \textbf{Layout Simplicity:} Misaligned text, images, or other elements.
\end{itemize}}
\subsection{Color Contrast}
\subsubsection{Round 1}
\changes{\begin{itemize}
\item \textbf{Poor Text Contrast:} Insufficient contrast between text and background.
\item \textbf{Color Scheme Issues:} Inconsistent or poorly chosen color schemes leading to contrast issues.
\item \textbf{Poor Background Contrast:} Background not providing enough contrast with the text or other elements.
\item \textbf{Background Clarity:} Background being too bright or too dark, affecting the overall design.
\item \textbf{Prominence of Elements:} Elements not being visually prominent due to poor contrast.
\item \textbf{Icon and Label Visibility:} Icons or labels being hard to see or understand.
\end{itemize}
\subsubsection{Round 2}
\begin{itemize}
\item \textbf{Poor Text Contrast:} Insufficient contrast between text and background.
\item \textbf{Icon Contrast Issues:} Icons being hard to see due to poor contrast.
\item \textbf{Poor Button Contrast:} Insufficient contrast between button and background, or poor readability of button label due to low contrast.
\item \textbf{Poor Element Contrast:} Poor contrast with background for other UI elements, such as images, sliders, etc.
\end{itemize}}
\subsection{Text Readability}
\subsubsection{Round 1}
\changes{\begin{itemize}
\item \textbf{Font Size:} The font size is too small or difficult to read.
\item \textbf{Font Weight:} The need to increase or change the font weight for better readability.
\item \textbf{Font Style:} Changes to the font style to improve legibility.
\item \textbf{Hierarchy and Emphasis:} Lack of visual hierarchy or emphasis in the text elements.
\item \textbf{Consistency:} Inconsistent font sizes, styles, or weights across the design.
\item \textbf{Content Clarity:} Need for clear and concise content to improve readability and user understanding.
\item \textbf{Visual Prominence:} Important text elements should be more visually prominent.
\item \textbf{Element Overlap:} Issues where text elements overlap with other UI elements.
\item \textbf{Redundancy and Density:} Text is too dense or contains redundant information that could be simplified.
\end{itemize}}
\subsubsection{Round 2}
\changes{\begin{itemize}
    \item \textbf{Font Size:} The font size is too small or difficult to read.
    \item \textbf{Font Weight:} The need to increase or change the font weight for better
    \item \textbf{Text Density:} Text is too dense, which could cause overlaps with other elements, or contains redundant information that could be simplified.
\end{itemize}}
\subsection{Usability of Buttons}
\subsubsection{Round 1}
\changes{\begin{itemize}
    \item \textbf{Button Placement:} Confusion due to button location.
    \item \textbf{Button Visual Design:} Need for visual differentiation using color, size, and style.
    \item \textbf{Spacing and Alignment:} Uneven spacing between buttons.
    \item \textbf{Button Recognition:} Issues with buttons not looking like interactive elements.
    \item \textbf{Button Size:} Buttons being too small or too large.
    \item \textbf{Call-to-Action Clarity:} Buttons or text labels not clearly indicating their function.
    \item \textbf{Button Style:} Busy or outdated button styles.
    \item \textbf{Color Scheme:} Mismatched button colors and overall design.
    \item \textbf{User Feedback and Affordances:} Lack of visual feedback on button presses.
    \item \textbf{Button and Text Alignment:} Misaligned buttons and text fields.
    \item \textbf{Hierarchy and Flow:} Misplaced primary action buttons disrupting visual flow.
    \item \textbf{Button Spacing:} Buttons placed too close together.
    \item \textbf{Button Emphasis:} Primary actions not visually emphasized.
    \item \textbf{User Control Elements:} Missing user control buttons.
    \item \textbf{Background Contrast:} Low contrast between background and buttons.
    \item \textbf{Button Consistency:} Inconsistent button designs.
    \item \textbf{Button Visual Hierarchy:} Important buttons are not visually prominent.
\end{itemize}}
\subsubsection{Round 2}
\changes{\begin{itemize}
    \item \textbf{Button Visual Design:} All aspects of the visual design of the button or button group for improved usability.
    \item \textbf{Button Clarity:} Clarity of the purpose of the button through text labels, etc
    \item \textbf{Addition of Buttons:} The addition of buttons to simplify tasks and increase user control.
\end{itemize}}
\subsection{Learnability}
\subsubsection{Round 1}
\changes{\begin{itemize}
    \item \textbf{Inappropriate Icon:} The icon used does not match the intended message or function.
    \item \textbf{Unclear Icon Meaning:} Icons do not clearly convey their meaning or function.
    \item \textbf{Missing Interactivity Indication:} Elements that should indicate interactivity (e.g., checkboxes, buttons) are missing or unclear.
    \item \textbf{Missing Functionality:} Functional elements (e.g., buttons) are missing, making the app less intuitive to use.
    \item \textbf{Unclear UI Region:} Purpose of an element group or region in the UI is not clear.
    \item \textbf{Poor Visual Hierarchy:} Elements lack a clear visual hierarchy, making the interface confusing.
    \item \textbf{Unrealistic Icon:} Icons do not represent real-world objects or concepts.
    \item \textbf{Unclear UI Purpose:} The purpose of the UI screen is not intuitive.
    \item \textbf{Missing Placeholder:} Placeholder text is missing, making it hard for users to figure out what type of information the field asks for.
    \item \textbf{Unclear Element Functionality:} Elements do not clearly indicate their functionality.
    \item \textbf{Missing Labels:} Elements are missing labels, causing confusion.
\end{itemize}}
\subsubsection{Round 2}
\changes{\begin{itemize}
    \item \textbf{Unclear Functionality:} Icons or other UI elements do not clearly indicate their functionality
    \item \textbf{Unclear Purpose:} The purpose of regions in the UI or the entire UI screen is unclear
    \item \textbf{Unclear Labels:} The text labels do not match the icon or UI element, or do not clearly explain the purpose of the UI region or screen.
\end{itemize}}

\end{document}